\documentclass[letter]{aa}  
\usepackage{ulem}
\usepackage{color}
\definecolor{mygray}{gray}{0.6}
\usepackage{bm}

\usepackage{graphicx}
\usepackage{float}

\usepackage[varg]{txfonts}
\usepackage{natbib}	
\usepackage{hyperref}
\graphicspath{{./plots/}}
\begin{document}


   \title{Radial and vertical dust transport inhibit refractory carbon depletion in protoplanetary disks}

   \author{L. Klarmann  \inst{1}
          \and
          C. W. Ormel  \inst{1}
          \and
          C. Dominik  \inst{1}
                    }

   \institute {Astronomical Institute Anton Pannekoek, University of Amsterdam, Science Park 904, 1098 XH Amsterdam, The Netherlands\\
              \email{l.a.klarmann@uva.nl}}

    \date{Received June 26, 2018; Accepted September 5, 2018}

\abstract
{The Earth is strongly depleted in carbon compared to the dust in the ISM, implying efficient removal of refractory carbon before parent body formation. It has been argued that grains get rid of their carbon through oxidation and photolysis in the exposed upper disk layers.} 
{We assess the efficacy of these C-removal mechanisms accounting for the vertical and radial transport of grains.}
{We obtain the carbon and carbon free mass budget of solids by solving two 1D advection-diffusion equations, accounting for the dust grain size distribution and radial transport. The carbon removal acts on the fraction of the grains that are in the exposed layer and requires efficient vertical transport.}
{In models without radial transport, oxidation and photolysis can destroy most of the refractory carbon in terrestrial planet formation region. But it only reaches the observed depletion levels for extreme parameter combinations and requires that parent body formation was delayed by 1\,Myr. Adding radial transport of solids prevents the depletion entirely, leaving refractory carbon equally distributed throughout the disk.}
{It is unlikely that the observed carbon depletion can ultimately be attributed to mechanisms operating on small grains in the disk surface layers. Other mechanisms need to be studied, for example flash heating events or FU Ori outbursts in order to remove carbon quickly and deeply. However, a sustained drift barrier or strongly reduced radial grain mobility are necessary to prevent replenishment of carbon from the outer disk.}

   
\titlerunning{Radial and vertical dust transport inhibit refractory carbon depletion in PPDs}
\maketitle

\section{Introduction}

The Earth is significantly depleted in carbon \citep{Allegre01}. Its silicon to carbon ratio is a factor of 10$^{-4}$ lower than in the Sun \citep{Grevesse10} or in the ISM, the base material it formed from \citep{Bergin15}. The picture is different for objects formed further out in the solar nebula. The silicon to carbon ratio of carbonaceous chondrites is only a factor 100 lower than in the ISM \citep{Wasson88}, and many comets are not carbon depleted at all \citep{Wooden08}. This seems to show a gradient of carbon depletion in the solar system, with depletion getting stronger for objects that formed closer to the Sun \citep[e.g.][]{Pontoppidan14, Lee10,Geiss87}. 

However, more than half of the carbon in the ISM is expected to be refractory material \citep{Zubko04}. Several processes have been suggested to remove this refractory carbon from the inner regions of a protoplanetary disks. \citet{Gail17} investigate the destruction of refractory carbon species within their radial transport models \citep[see also][]{Gail01}, but find that oxidation in the disk mid-plane via OH molecules \citep{Finocchi97} is not sufficient to deplete the inner disk region of carbon. \citet{Anderson10} employ carbon oxidation via atomic oxygen in the hot, upper disk layer \citep[following][]{Lee10} and the photolysis of carbon grains directly via UV photons \citep{Alata14}. In the inner disk regions they reach a carbon depletion comparable to the Earth, but only if all refractory material is in small grains and without taking into account radial dust transport.

In this work, we investigate how the presence of large grains and vertical and radial dust transport influence the refractory carbon in a protoplanetary disk, and the viability of depleting the inner disk region via oxidation and photolysis in the upper disk layers.

\section{Model}

\subsection{Disk model}
For the star we take $M_{\star}=1\,\mathrm{M_{\odot}}$, $L_{\star}=1\,\mathrm{L_{\odot}}$ and the UV field is set to $L_{\mathrm{UV}}=0.01L_{\star}$. The total disk mass is set to 0.039$\,\mathrm{M_{\odot}}$. We adopt a dust-to-gas ratio of 0.01, similar to \citet{Anderson10} and \citet{Kamp17}. The disk surface density follows a power-law profile with $\Sigma\propto r^{-1}$ and is exponentially cut-off at 200\,au.
The carbon mass fraction $f_c(r)$ is defined as the ratio of the solid carbon surface density and the total dust surface density, $f_{\mathrm{c}}=\Sigma_{\mathrm{c}}/\Sigma_{\mathrm{tot}}$. We assume an initial carbon-to-hydrogen abundance of 2$\cdot10^{-4}$, which agrees with solar \citep{Asplund09} and ISM abundances \citep{Jenkins09}. Like \citet{Anderson10}, we divide carbon equally between volatiles and refractory grains \citep[see also][]{Zubko04}. This leads to an initial refractory carbon mass fraction of $f_{\mathrm{c}}\approx 0.25$.

We use a 1+1\,D approach to describe the dust movement and composition. The radial dust transport, together with grain growth and fragmentation, is modelled using the \texttt{twopoppy} code by \citet{Birnstiel12, Birnstiel15} and using a fragmentation velocity $v_{\mathrm{f}}=10\, \mathrm{m\, s^{-1}}$ throughout the disk (see App.~\ref{app:twopoppy}).

\begin{figure}
    \centering
    \includegraphics[width=0.99\columnwidth,trim={3.5cm 17.5cm 6.5cm 3.4cm},clip]{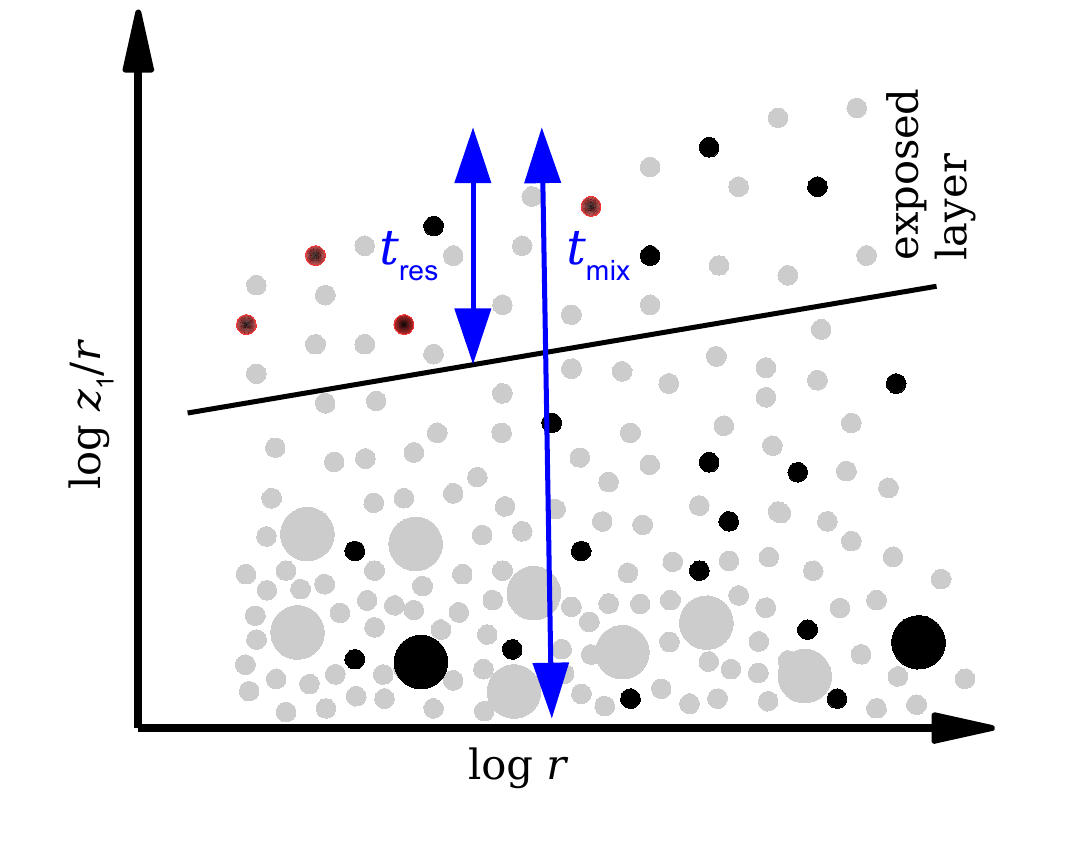}
    \caption{Height of the exposed layer divided by the disk radius $z_1/r$ against disk radius $r$. Silicate grains are plotted in gray, carbon grains in black. Large grains are settled close to the midplane. In the exposed layer above $z_1$, carbon grains can be oxidised, indicated in red. The range of vertical dust movement shown by blue arrows. Grains stay for $t_{\mathrm{res}}$ in the exposed layer, and within $t_{\mathrm{mix}}$, material from the exposed layer is well-mixed with material from the midplane.}
    \label{fig:cartoon}
\end{figure}

We refer to the surface layer of the disk where the refractory carbon gets destroyed as the \textit{exposed layer}, because that layer is exposed to UV photons from the star. This is the layer where carbon reacts with free oxygen and where most of the photolysis occurs. We denote the vertical coordinate of the exposed layer as $z_{1}$ and its dust surface density as $\Sigma^\ast$. 

Figure~\ref{fig:cartoon} shows the model setup, depicting the exposed layer above a height $z_1$ as a function of the disk radius $r$. The blue arrows indicate the dust mixing between the midplane and the exposed layer ($t_{\rm mix}$) and the removal and replacement of dust in the exposed layer ($t_{\rm res}$) (see Sec.~\ref{sub:vertical}). In the exposed layer, carbon can be destroyed via oxidation or photolysis, indicated in red.

\subsection{Calculation of the location of the exposed layer}
\label{sub:calcz1}
The exposed layer is the layer that can be reached by stellar photons. Determining the height $z_1$ of the exposed layer comes down to determining the height where small grains still coupled to the gas at that height do provide the required optical depth. Because of the flaring geometry of the disk, a radial optical depth $\tau_r=1$ corresponds to a vertical depth $\tau_z=\Phi$ where $\Phi=0.05$ is the disk flaring angle. The value of $z_1$ depends on the total dust surface density $\Sigma_{\mathrm{tot}}$, the grain size distribution, the grain opacity $\kappa$, and grain settling. We assume that the grains in the exposed layer are in the Rayleigh regime and take $\kappa=\kappa_0=2\cdot10^4\,\mathrm{cm^2/g}$ independent of grain radius $s$.\footnote{We use $\kappa_{0} = 2\cdot10^4\,\mathrm{g/cm^2}$, which corresponds to a grain size distribution up to 0.1\,$\mathrm{\mu}$m with $f_{\mathrm{c}}$=0.15 \citep{Min16}. A correct opacity treatment would take the local grainsize distribution and $f_\mathrm{c}$ into account, as well as icy grains in the outer disk. This could change $\kappa_0$ by a factor of about five, an effect we explore in a small parameter study.} The surface density $\Sigma^\ast$ of the exposed layer then follows, $\Sigma^\ast = \Phi/\kappa_{\mathrm{0}}$. For simplicity, we assume that large grains in the optically geometrical limit, $s>s_{\mathrm{geo}}=\lambda/2\pi \approx0.1\,\mathrm{\mu}$m for $\lambda=0.55\,\mathrm{\mu}$m, do not contribute to the optical opacity. Then,
\begin{equation}
    \tau_z(s,z) = f_{\leq s} f_{\geq z} \Sigma \kappa_0
\end{equation}
where $f_{\leq s} = (s/s_{\mathrm{max}})^{4-p}$ is the fraction by mass of grains smaller than radius $s$, assuming a power-law size distribution with exponent with $p=3.5$ and maximum grain radius $s_{\mathrm{max}}$ determined by drift and fragmentation (see App.~\ref{app:twopoppy}).
Similarly $f_{\geq z}$ is the fraction of the surface density above height $z$ 
\begin{equation}
    f_{\ge z}(s,z) = \frac{1}{2}\textrm{erfc}\left( \frac{z}{\sqrt{2}h_{\mathrm{gr}}(s,z)} \right).
\end{equation}
In calculating this fraction we use the dust scale height $h_{\mathrm{gr}}$ based on the \textit{local} Stokes number St corresponding to $z$ and the turbulent $\alpha$ parameter, $h_{\mathrm{gr}} = H \sqrt{\alpha/\left(\alpha +\mathrm{St}(z,s)\right)}$.
Hence $f_{\leq s}$ increases with $s$ while $f_{\geq z}$ (for a fixed $z$) decreases with $s$.

\begin{figure}
\centering
    \includegraphics[width=1.0\columnwidth]{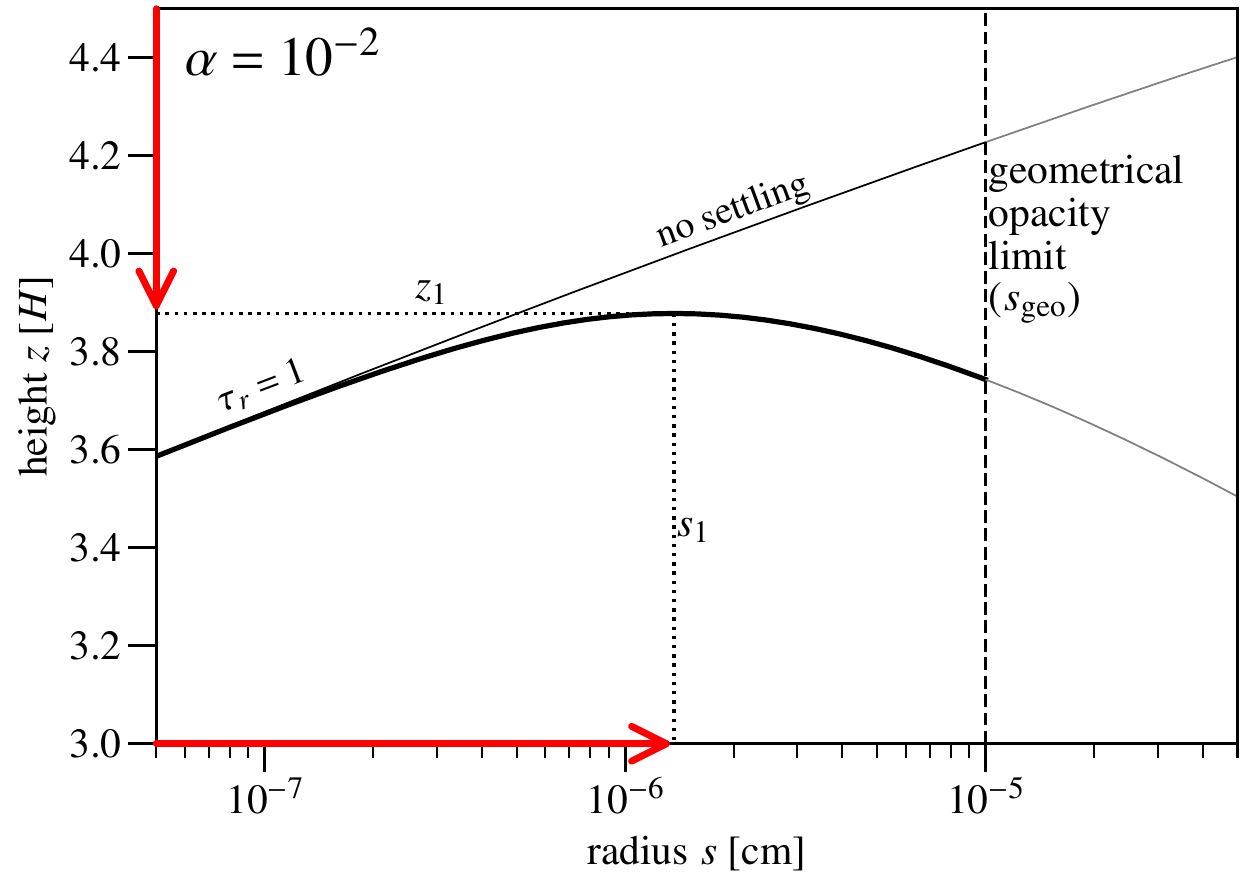}
    \caption{Height $z$ against grain radius $s$ where $\tau_r=1$ at $r=1\,$au. The point where the z coordinate of $\tau_r$ peaks defines the height of the exposed layer $z_1$ and the radius $s_1$ up to which grains contribute to the opacity in the exposed layer (red arrows). Grains larger than $s_1$ are too settled to contribute. The size $s_{\mathrm{geo}}$ is an upper limit for $s_1$ as the opacity decreases with larger grain sizes.
}
      \label{fig:z1-new}
\end{figure}

Figure~\ref{fig:z1-new} illustrates these points, plotting the height $z$ where $\tau_{r}=1$ as function of grain radius $s$. Grains contribute to the opacity build up in the exposed layer up to a size $s_1$. Ignoring their settling, all grains would contribute to the exposed layer which would therefore lie very high in the disk. In reality settling causes the largest grains to drop out of the exposed layer. We identify the point where $\tau_r=1$ peaks as the size $s_1$ and the height $z_1$. \footnote{In our definition of $f_{\geq z}(s,z)$, it is assumed that all grains smaller than $s$ settle to the same height as grains with radius $s$, which is the reason why $f_{\geq z}(s,z)$ eventually decreases with size $s$. Although, this would be incorrect in a cumulative sense (extending the distribution to include larger grains should not decrease the mass fraction) it here simply serves the point of finding the size where grains decouple.}

\begin{figure*}[t]
\sidecaption
\includegraphics[width=0.7\textwidth]{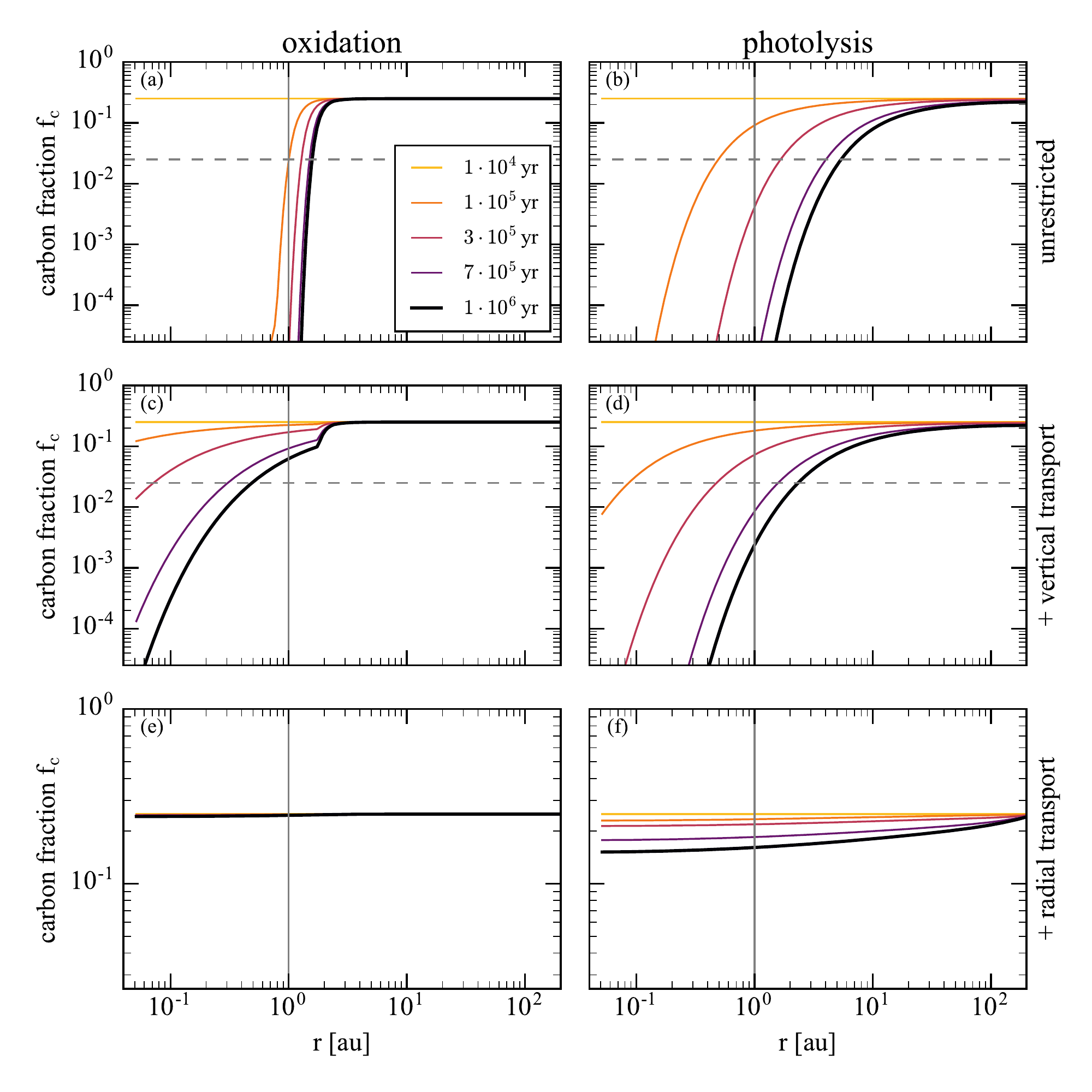}
\caption{All panels show the carbon fraction $f_{\mathrm{c}}=\Sigma_\mathrm{c}/\Sigma_\mathrm{tot}$ as function of disk radius $r$. Coloured lines show the time evolution of the carbon fraction $f_{\mathrm{c}}$ in the fiducial model from $10^4$\,yr to $10^6$\,yr. The vertical line indicates Earth's position. The horizontal dashed line shows depletion by a factor ten. Left: carbon removal by oxidation. Right: carbon removal by photolysis. Top panels: the oxidation (photolysis) $f_c\Sigma^{\ast} /t_{\mathrm{ox}}$ ($f_c\Sigma_{\mathrm{ph}}^{\ast} /t_{\mathrm{ph}}$) is applied to the carbon in the exposed layer unrestricted by vertical or radial transport. Middle panels: vertical dust transport is taken into account as described in Eq.~\ref{eq:sink} (Eq.~\ref{eq:sink-ph}). Bottom panels: vertical and radial dust transport as described in App.~\ref{app:twopoppy} are included. The scale of the y-axis has been changed because of the low level of carbon depletion in these models.}
\label{fig:fiducial}
\end{figure*}

\subsection{Carbon removal}
Carbon removal in the exposed layer occurs via oxidation or photolysis (see App.~\ref{app:C-removal}). The oxidation time for one carbon grain in the exposed layer is
\begin{equation}
    \label{eq:tox}
t_{\mathrm{ox}}
= \frac{4}{3}\frac{s_1 \rho_{\mathrm{c}}}{n_{\mathrm{ox}} v_{\mathrm{ox}} Y_{\mathrm{ox}} m_{\mathrm{c}}}
\end{equation}
where $\rho_{\mathrm{c}}=2\,$g/cm$^3$ is the specific density of the carbon grain material, $m_{\mathrm{c}}$ the mass of a carbon atom, $n_{\mathrm{ox}}$ the number density of oxygen atoms at $z_1$, $v_{\mathrm{ox}}$ the oxygen thermal velocity and $Y_\mathrm{ox}$ the yield of the oxidation \citep{Draine79}.
The photolysis rate is
\begin{equation}
\label{eq:dS/dt_ph}
\frac{d\Sigma_{\mathrm{c}}}{dt} = \Phi F_{\mathrm{UV}}  Y_\mathrm{ph}   m_{\mathrm{c}} f_{\mathrm{c}}\,\;,
\end{equation}
where $F_{\mathrm{UV}}$ is the UV flux and $m_{\mathrm{c}}$ the mass of a carbon atom. The flaring angle $\Phi$ corrects for the fact that the UV photons do not hit the disk surface perpendicular. The yield is $Y_\mathrm{ph}=8\cdot 10^{-4}$ \citep{Alata14, Alata15}.
A more detailed description of these processes can be found in \citet{Lee10} and \citet{Anderson10} and in Appendix~\ref{app:C-removal}.

\begin{table*}[htb]
\centering
\begin{tabular}{llll}
\hline 
variation & value & \multicolumn{2}{c}{$f_{\mathrm{c}}$ at 1\,au after 1\,Myr} \\ 
 &  &only vertical transport& vertical + radial transport \\ 
\hline 
\hline
fiducial    &   & 1.7$\cdot10^{-3}$ & 0.16 \\ 
higher UV-flux  & $L_\mathrm{UV}=0.1L_\star$ & 5.2$\cdot10^{-5}$ & 0.11 \\ 
lower opacity   & $\kappa_{0}=4\cdot10^3\,\mathrm{cm^2/g}$ & $6.4\cdot 10^{-11}$ & 0.052 \\ 
larger opacity  & $\kappa_{0}=1\cdot10^5\,\mathrm{cm^2/g}$ & 0.10 & 0.23 \\ 
lower turbulence & $\alpha=10^{-3}$ & 0.015 & 0.24 \\ 
less small grains  & $p=3$ & 0.061 & 0.23 \\ 
lower fragmentation velocity & $v_{\mathrm{f}}=3\,$m/s & 4.0$\cdot 10^{-8}$ & 0.12\\
\hline 
\end{tabular}
\caption{Carbon fraction $f_{\mathrm{c}} = \Sigma_c/\Sigma_\mathrm{tot}$ after 1\,Myr at 1\,au for models limited by vertical transport and models limited by vertical and radial transport. Initial carbon fraction is $f_{\mathrm{c}}=0.25$. Only one parameter is varied with respect to the fiducial model.} 
\label{tab:parameterspace}
\end{table*}

\subsection{Effects of vertical dust transport on removal rates}
\label{sub:vertical}
The efficacy of carbon removal is limited by the ability to vertical transport (cycle) the dust. There are two important timescales regarding to the vertical motions of grains. The first is the overall mixing time $t_\mathrm{mix}$, indicating on what timescale material from the midplane and the exposed layer become well-mixed. We obtain $t_\mathrm{mix}$ from the turbulent diffusivity and the gas scale height $H$:
\begin{equation}
 t_{\mathrm{mix}}
 =\frac{H^2}{\nu_{\mathrm{t}}}
 =\frac{1}{\Omega \alpha}
 = \mathrm{100\,yr} \left( \frac{\alpha}{10^{-2}} \right)^{-1} \left( \frac{r}{\mathrm{au}} \right)^{3/2}
 \end{equation}
 where the turbulent diffusivity is assumed equal to the gas viscosity $\nu_{\mathrm{t}}$ and is parametrised using the $\alpha$ prescription ($\nu_t=\alpha H^2 \Omega$; \citealt{Shakura73}) with $\Omega$ the Keplerian frequency. After one mixing timescale, carbon-depleted material from the exposed layer and carbon-rich material from the midplane are well mixed.

The other relevant timescale is the residence time\footnote{\citet{Anderson10} define $t_{\mathrm{res}}$ differently, as the \textit{total} time a grain spends in the exposed layer over 1\,Myr.} of a grain in the exposed layer
\begin{equation}
t_{\mathrm{res}}
= \left( \frac{H}{z_{1}}\right)^2 \frac{1}{\Omega \alpha} 
= \mathrm{11\,yr}  \left( \frac{z_1}{3H} \right)^{-2} \left( \frac{\alpha}{10^{-2}} \right)^{-1} \left( \frac{r}{\mathrm{au}} \right)^{3/2}.
\end{equation}
This time is shorter than $t_\mathrm{mix}$ because the local pressure scale height at height $z\gg H$ is given by $H^2/z$.
A long $t_\mathrm{res}$ would limit carbon destruction, because the exposed layer will become depleted in refractory carbon. In that case no carbon will be burned, because carbon-free solids build up the opacity in the exposed layer. Hence, carbon removal becomes inefficient when $t_\mathrm{res}$ is longer than the time to burn a single grain $t_\mathrm{ox}$ (see Eq.~(\ref{eq:tox})). 

Accounting for these vertical transport effects, we obtain a carbon destruction rate of:
\begin{equation}
    \label{eq:sink}
\frac{d\Sigma_{\mathrm{c}}}{dt} = 
2f_{\mathrm{c}}\cdot \mathrm{min}\left( 
\frac{\Sigma^{\ast}}{t_{\mathrm{ox}}},  
\frac{\Sigma^{\ast}}{t_{\mathrm{res}}},
\frac{f_{\leq s_1}\Sigma_\mathrm{tot}}{t_{\mathrm{mix}}} \right)
\end{equation}
where the factor 2 accounts for the two sides of the disk and $\Sigma_{\mathrm{c}}$ is the surface density of carbon grains with $s<s_1$. This expression applies to C-burning; in App.~\ref{app:photolysis} a similar expression is derived for photolysis.


\newcommand{\eq}[1]{Eq.\,(\ref{eq:#1})}
\newcommand{\fg}[1]{Fig.\,\ref{fig:#1}}

\section{Results}
Our results are presented in Fig.~\ref{fig:fiducial} for the oxidation (left) and the photolysis (right) models. 

\subsection{Unrestricted models}
The upper row plots present the carbon fraction $f_{\mathrm{c}}$ without including any transport-limiting factors, i.e., by using only the first term in \eq{sink} ($f_c\Sigma^{\ast} /t_{\mathrm{ox}}$). As can be seen, oxidation depletes carbon by a factor of $10^{-4}$ out to 1.1\,au. Beyond this point carbon burning is essentially shut off, because of the exponential dependence of the oxidation yield $Y_{\mathrm{ox}}$ on temperature. The photolysis rate, on the other hand, does not depend on temperature ($F_\mathrm{UV}\propto r^{-2}$ but $Y_\mathrm{ph}$ is constant). After 1\,Myr, the disk is depleted by a factor of $10^{-4}$ out to 1.3\,au and by a factor of 0.1 out to 7\,au.

\subsection{Vertical dust transport}
Accounting for vertical transport effects -- i.e., including all three regimes in Eq.~(\ref{eq:sink}) -- we see that oxidation (Fig.~\ref{fig:fiducial}c) and photolysis (Fig.~\ref{fig:fiducial}d) become less effective. Carbon oxidation inward of 1.3\,au becomes now limited by the residence time $t_\mathrm{res}$. Grains that make it into the exposed layer burn their carbon atoms completely, rendering the overall burning inefficient. Similarly, the photolysis rate equals $f_{\mathrm{c}}\Sigma^{\ast}_{\mathrm{ph}}/t_{\mathrm{res-ph}}$ everywhere. The photolysis rate tends to be larger than the oxidation rate, because the UV photons penetrate more deeply, resulting in a larger exposed layer (see App.~\ref{app:photolysis}). However, the photolysis run is just short of reaching depletion levels of $10^{-4}$ at 1\,au.

\subsection{Vertical and radial dust transport}
Accounting in addition for radial transport (bottom panels of Fig.~\ref{fig:fiducial}) further reduces the efficacy of carbon destruction. Carbon-rich solids from the outer disk simply drift into the inner region to replenish any carbon depleted material. The carbon destruction becomes drift-limited: carbon will only be depleted when the local destruction time ($t_{\mathrm{destr}} = \Sigma_{\mathrm{c}}/(d\Sigma_{\mathrm{c}}/dt)$) becomes shorter than the drift timescale $t_{\mathrm{drift}}$ of the (mass-dominating) pebbles. Since $t_{\mathrm{drift}}$ tends to be rather short (200\,yr at 1\,au) depletion is minimal with little variation throughout the disk.

\subsection{Parameter variation}
 In Table~\ref{tab:parameterspace} we list the results from additional photolysis runs, quantifying the level of  carbon depletion at 1\,au after 1\,Myr. A higher opacity, a lower $\alpha$ or a shallower grain size distribution ($p=3$; fewer small grains) only reduce the carbon destruction, because $\Sigma^{\ast}$ becomes lower or $t_\mathrm{res}$ increases. For the runs without radial transport, a stronger UV field enhances the depletion as the UV photons penetrate deeper.  It can be argued that the low $\kappa_{0}$ run is more appropriate for the opacity in the exposed layer, when the grains lose most of their carbon. This will increase $\Sigma^{\ast}$, and completely remove all carbon interior to 1\,au -- but only when there is no radial replenishment. Similarly, reducing the fragmentation velocity to 3\,m/s leads to more small grains and therefore a stronger depletion at 1\,au in the case of only vertical transport. When radial transport is included, the carbon fraction is only reduced by a factor of two. Since the fragmentation velocity outside of the snowline is expected to be much higher than 3\,m/s, a more realistic case where the fragmentation velocity depends on the disk radius would lead to even less carbon depletion.

In general, the replenishment of carbon via radial transport renders the depletion independent of the adopted parameters.

\section{Discussion and Conclusion}

Our findings regarding unrestricted C-burning are in line with the study of \citet{Anderson10}. They, too, find that photolysis is the more significant C-depletion mechanism and that the inner disk can become devoid in carbon. Like us, \citet{Anderson10} account for the limited \textit{total} time grains spend in the exposed layers. However, they have overestimated the removal by incorrectly assuming that each stay in the exposed layer is short enough to ensure a continuous supply of carbon in the exposed layer. Instead, we found (even in the case of high turbulence) that carbon-depleted grains stay longer in the exposed layer than it takes to remove their carbon. This makes both oxidation and photolysis inefficient. Furthermore, adding radial dust transport to the model makes it impossible to deplete the inner disk of carbon even under extreme assumptions about the UV field of the early Sun or the grain opacities. 

It is unlikely that other C-destruction mechanisms can change this outcome. Adding oxidation of carbon by OH in the midplane does not increase the carbon depletion sufficiently \citep{Gail17}. For high accretion rates, the midplane region can be heated to roughly 1500\,K out to 2\,au \citep{Min11}. However, these accretion rates must then be sustained over a significant time to allow the Earth's building blocks to form.

Therefore, we conclude that the only way to ensure the C depletion factors as observed in the Solar System is to invoke an early hot or intense inner disk environment to ensure rapid C-destruction before parent body formation. This needs to happen in combination with a sustained barrier for drift to prevent C-replenishment, for example by the formation of a giant planet. FU Orionis events can lead to inner disk temperatures of several thousand Kelvin \citep{Hartmann96,Zhu07} over several decades. The composition of Chondrules indicate that several flash heating events happened in the solar nebula \citep{Ciesla05, Jones00}, reaching temperatures of around 2000\,K nearly instantly and cooling again within days. To prevent the fast replenishment of carbon, these events need to happen at a high frequency, comparable to the drift timescale. Events at a lower rate can also cause a sustained refractory carbon depletion in the inner disk region, but this requires radial grain mobility to be strongly reduced or halted.

\paragraph{\textit{Acknowledgements}}The authors thank the referee for valuable feedback. The authors thank Inga Kamp for sharing her \texttt{ProDiMo} model, Christian Rab for sharing \texttt{prodimopy}, Michiel Min for sharing \texttt{OpacityTool} and Til Birnstiel for sharing \texttt{twopoppy}. We thank the above-mentioned and Rens Waters and Kaustubh Hakim for helpful discussions and Sebastiaan Krijt for his thoughtful comments on the manuscript. L.K. is supported by a grant from NOVA. C.W.O. acknowledges funding by the Netherlands Organization for Scientific Research (NWO; VIDI project 639.042.422). C.D. acknowledges funding by the NWO, project number 614.001.552.

\bibliography{cb_references}

\begin{thebibliography}{31}
\expandafter\ifx\csname natexlab\endcsname\relax\def\natexlab#1{#1}\fi

\bibitem[{{Alata} {et~al.}(2014){Alata}, {Cruz-Diaz}, {Mu{\~n}oz Caro}, \&
  {Dartois}}]{Alata14}
{Alata}, I., {Cruz-Diaz}, G.~A., {Mu{\~n}oz Caro}, G.~M., \& {Dartois}, E.
  2014, \aap, 569, A119

\bibitem[{{Alata} {et~al.}(2015){Alata}, {Jallat}, {Gavilan}, {Chabot},
  {Cruz-Diaz}, {Munoz Caro}, {B{\'e}roff}, \& {Dartois}}]{Alata15}
{Alata}, I., {Jallat}, A., {Gavilan}, L., {et~al.} 2015, \aap, 584, A123

\bibitem[{{All{\`e}gre} {et~al.}(2001){All{\`e}gre}, {Manh{\`e}s}, \&
  {Lewin}}]{Allegre01}
{All{\`e}gre}, C., {Manh{\`e}s}, G., \& {Lewin}, {\'E}. 2001, Earth and
  Planetary Science Letters, 185, 49

\bibitem[{{Anderson} {et~al.}(2017){Anderson}, {Bergin}, {Blake}, {Ciesla},
  {Visser}, \& {Lee}}]{Anderson10}
{Anderson}, D.~E., {Bergin}, E.~A., {Blake}, G.~A., {et~al.} 2017, \apj, 845,
  13

\bibitem[{{Asplund} {et~al.}(2009){Asplund}, {Grevesse}, {Sauval}, \&
  {Scott}}]{Asplund09}
{Asplund}, M., {Grevesse}, N., {Sauval}, A.~J., \& {Scott}, P. 2009, \araa, 47,
  481

\bibitem[{{Bergin} {et~al.}(2015){Bergin}, {Blake}, {Ciesla}, {Hirschmann}, \&
  {Li}}]{Bergin15}
{Bergin}, E.~A., {Blake}, G.~A., {Ciesla}, F., {Hirschmann}, M.~M., \& {Li}, J.
  2015, Proceedings of the National Academy of Science, 112, 8965

\bibitem[{{Birnstiel} {et~al.}(2015){Birnstiel}, {Andrews}, {Pinilla}, \&
  {Kama}}]{Birnstiel15}
{Birnstiel}, T., {Andrews}, S.~M., {Pinilla}, P., \& {Kama}, M. 2015, \apjl,
  813, L14

\bibitem[{{Birnstiel} {et~al.}(2012){Birnstiel}, {Klahr}, \&
  {Ercolano}}]{Birnstiel12}
{Birnstiel}, T., {Klahr}, H., \& {Ercolano}, B. 2012, \aap, 539, A148

\bibitem[{{Ciesla}(2005)}]{Ciesla05}
{Ciesla}, F.~J. 2005, in Astronomical Society of the Pacific Conference Series,
  Vol. 341, Chondrites and the Protoplanetary Disk, ed. A.~N. {Krot}, E.~R.~D.
  {Scott}, \& B.~{Reipurth}, 811

\bibitem[{{Draine}(1979)}]{Draine79}
{Draine}, B.~T. 1979, \apj, 230, 106

\bibitem[{{Fedele} {et~al.}(2016){Fedele}, {van Dishoeck}, {Kama}, {Bruderer},
  \& {Hogerheijde}}]{Fedele16}
{Fedele}, D., {van Dishoeck}, E.~F., {Kama}, M., {Bruderer}, S., \&
  {Hogerheijde}, M.~R. 2016, \aap, 591, A95

\bibitem[{{Finocchi} {et~al.}(1997){Finocchi}, {Gail}, \&
  {Duschl}}]{Finocchi97}
{Finocchi}, F., {Gail}, H.~P., \& {Duschl}, W.~J. 1997, \aap, 325, 1264

\bibitem[{{Gail}(2001)}]{Gail01}
{Gail}, H.~P. 2001, \aap, 378, 192

\bibitem[{{Gail} \& {Trieloff}(2017)}]{Gail17}
{Gail}, H.-P. \& {Trieloff}, M. 2017, \aap, 606, A16

\bibitem[{{Geiss}(1987)}]{Geiss87}
{Geiss}, J. 1987, \aap, 187, 859

\bibitem[{{Grevesse} {et~al.}(2010){Grevesse}, {Asplund}, {Sauval}, \&
  {Scott}}]{Grevesse10}
{Grevesse}, N., {Asplund}, M., {Sauval}, A.~J., \& {Scott}, P. 2010, \apss,
  328, 179

\bibitem[{{Hartmann} \& {Kenyon}(1996)}]{Hartmann96}
{Hartmann}, L. \& {Kenyon}, S.~J. 1996, \araa, 34, 207

\bibitem[{{Jenkins}(2009)}]{Jenkins09}
{Jenkins}, E.~B. 2009, \apj, 700, 1299

\bibitem[{{Jones} {et~al.}(2000){Jones}, {Lee}, {Connolly}, {Love}, \&
  {Shang}}]{Jones00}
{Jones}, R.~H., {Lee}, T., {Connolly}, Jr., H.~C., {Love}, S.~G., \& {Shang},
  H. 2000, Protostars and Planets IV, 927

\bibitem[{{Kamp} {et~al.}(2017){Kamp}, {Thi}, {Woitke}, {Rab}, {Bouma}, \&
  {M{\'e}nard}}]{Kamp17}
{Kamp}, I., {Thi}, W.-F., {Woitke}, P., {et~al.} 2017, \aap, 607, A41

\bibitem[{{Lee} {et~al.}(2010){Lee}, {Bergin}, \& {Nomura}}]{Lee10}
{Lee}, J.-E., {Bergin}, E.~A., \& {Nomura}, H. 2010, \apjl, 710, L21

\bibitem[{{Meijerink} {et~al.}(2012){Meijerink}, {Aresu}, {Kamp}, {Spaans},
  {Thi}, \& {Woitke}}]{Meijerink12}
{Meijerink}, R., {Aresu}, G., {Kamp}, I., {et~al.} 2012, \aap, 547, A68

\bibitem[{{Min} {et~al.}(2011){Min}, {Dullemond}, {Kama}, \& {Dominik}}]{Min11}
{Min}, M., {Dullemond}, C.~P., {Kama}, M., \& {Dominik}, C. 2011, \icarus, 212,
  416

\bibitem[{{Min} {et~al.}(2016){Min}, {Rab}, {Woitke}, {Dominik}, \&
  {M{\'e}nard}}]{Min16}
{Min}, M., {Rab}, C., {Woitke}, P., {Dominik}, C., \& {M{\'e}nard}, F. 2016,
  \aap, 585, A13

\bibitem[{{Pontoppidan} {et~al.}(2014){Pontoppidan}, {Salyk}, {Bergin},
  {Brittain}, {Marty}, {Mousis}, \& {{\"O}berg}}]{Pontoppidan14}
{Pontoppidan}, K.~M., {Salyk}, C., {Bergin}, E.~A., {et~al.} 2014, Protostars
  and Planets VI, 363

\bibitem[{{Shakura} \& {Sunyaev}(1973)}]{Shakura73}
{Shakura}, N.~I. \& {Sunyaev}, R.~A. 1973, \aap, 24, 337

\bibitem[{{van Zadelhoff} {et~al.}(2003){van Zadelhoff}, {Aikawa},
  {Hogerheijde}, \& {van Dishoeck}}]{vanZadelhoff03}
{van Zadelhoff}, G.-J., {Aikawa}, Y., {Hogerheijde}, M.~R., \& {van Dishoeck},
  E.~F. 2003, \aap, 397, 789

\bibitem[{{Wasson} \& {Kallemeyn}(1988)}]{Wasson88}
{Wasson}, J.~T. \& {Kallemeyn}, G.~W. 1988, Philosophical Transactions of the
  Royal Society of London Series A, 325, 535

\bibitem[{{Wooden}(2008)}]{Wooden08}
{Wooden}, D.~H. 2008, \ssr, 138, 75

\bibitem[{{Zhu} {et~al.}(2007){Zhu}, {Hartmann}, {Calvet}, {Hernandez},
  {Muzerolle}, \& {Tannirkulam}}]{Zhu07}
{Zhu}, Z., {Hartmann}, L., {Calvet}, N., {et~al.} 2007, \apj, 669, 483

\bibitem[{{Zubko} {et~al.}(2004){Zubko}, {Dwek}, \& {Arendt}}]{Zubko04}
{Zubko}, V., {Dwek}, E., \& {Arendt}, R.~G. 2004, The Astrophysical Journal
  Supplement Series, 152, 211

\end{thebibliography}

\Online

\begin{appendix}

\section{Carbon removal}
\label{app:C-removal}
\subsection{Oxidation}

The exposed layer is characterized by a steep vertical gradient in gas temperature. However, most of the dust in the exposed layer will be found just above $z_{1}$, corresponding to a radial optical depth of unity. Since the penetration of UV photons is also responsible for the heating of the gas, this location can be very well characterized by a single temperature. Based on previous observations and thermo-chemical disk modelling \citep{Fedele16,Kamp17}, we describe the gas temperature in the exposed layer by
\begin{equation}
    T_{\mathrm{g}}=T_{\mathrm{i}} \left(\frac{r}{r_{\mathrm{i}}}\right)^{-q}\,\;.
\end{equation}
with $T_{\mathrm{i}}=750\,$K, $r_{\mathrm{i}}=1\,$au and $q=0.6$. This temperature leads to a mean thermal velocity of oxygen atoms of
\begin{equation}
    v_{\mathrm{ox}}=\sqrt{\frac{8k_{\mathrm{B}}T_{\mathrm{g}}}{\pi m_{\mathrm{ox}}}}\,\;,
\end{equation}
where $k_{\mathrm{B}}$ is Boltzmann's constant and $m_{\mathrm{ox}}$ is the mass of an oxygen atom.

The gas number density at $z_1$ is then
\begin{equation}
   n_{\mathrm{g}}= \frac{\Sigma_{\mathrm{g}}}{\sqrt{2\pi}\mu m_{\mathrm{p}} H} \exp{\left(-\frac{z_1^2}{2H^2} \right)}\,\;, 
\end{equation}
where $\Sigma_{\mathrm{g}}$ is the gas surface density, $\mu=2.35$ the mean molecular weight and $m_{\mathrm{p}}$ the proton mass. We use the oxygen number density at $z_1$, $n_{\mathrm{ox}}=\epsilon n_{\mathrm{g}}$, with $\epsilon=10^{-4}$ as found in \texttt{ProDiMo} models by \citet{Meijerink12} and similar to the value of $\epsilon\approx2\cdot10^{-4}$ shown in \citet{Lee10}.

The probability of removing a carbon atom when a carbon grain is hit by an oxygen atom is given by the yield
\begin{equation}
    Y_{\mathrm{ox}}=A\exp{(-B/T_{\mathrm{g}})}\,\;, 
\end{equation}
with $A=2.3, B=2580$ for $T_{\mathrm{g}}<440\,$K and $A=170, B=4430$ for $T_{\mathrm{g}}>440\,$K \citep{Draine79}.

The rate at which carbon is removed from a single grain by oxidation is
\begin{equation}
\label{app:eq:kox}
    k_{\mathrm{ox}}=n_{\mathrm{ox}}v_{\mathrm{ox}}\sigma Y_\mathrm{ox}\,\;,
\end{equation}
where $\sigma=\pi s^2$ is the grain cross section. For grains with radius $s_1$ this leads to a carbon destruction time of:

\begin{equation}
    \label{app:eq:tox}
t_{\mathrm{ox}}
= \frac{ m_{\mathrm{gr}}}{m_{\mathrm{c}} k_{\mathrm{ox}}}
= \frac{4}{3}\frac{s_1 \rho_{\mathrm{c}}}{n_{\mathrm{ox}} v_{\mathrm{ox}} Y_{\mathrm{ox}} m_{\mathrm{c}}}
\end{equation}
with $m_{\mathrm{gr}}$ the mass of a carbon grain. Grains lose all their carbon when they reside for a time $t\gg t_{\mathrm{ox}}$ in the exposed layer. The change in grain radius during oxidation is not taken into account in our model.

\subsection{Photolysis}
\label{app:photolysis}
In the case of photolysis by UV photons, a fraction $f_{\mathrm{c}}Y_\mathrm{ph}$ of the absorbed photons will remove a carbon atom directly, resulting in a destruction rate of:
\begin{equation}
\label{app:eq:dS/dt_ph}
\left(\frac{d\Sigma_c}{dt}\right)_\mathrm{ph-unrestricted} = \Phi F_{\mathrm{UV}}  Y_\mathrm{ph}   m_{\mathrm{c}} f_{\mathrm{c}}
\end{equation}

where $F_{\mathrm{UV}}$ is the UV field. The flaring angle $\Phi$ corrects for the fact that the UV photons do not hit the disk surface perpendicular. The yield is $Y_\mathrm{ph}=8\cdot 10^{-4}$ \citep{Alata14, Alata15, Anderson10}.

Analogous to oxidation, photolysis can also be limited by the residence timescale. However, in the case of photolysis UV photons can reach disk layers below the optical $\tau_{\mathrm{r}}=1$ line, due to forward scattering into the disk \citep{vanZadelhoff03}. We calculate the height $z_1$ of the layer exposed to UV radiation by equating $t_{\mathrm{res}}$, the residence time, with $t_{\mathrm{ph}}$, the time to destroy a carbon grain entirely by photolysis:

\begin{equation}
\label{z1_phot}
\left( \frac{H}{z_{1}}\right)^2 \frac{1}{\Omega \alpha}  = \frac{4}{3} \frac{s_1 \rho_{\mathrm{c}}}{F_{\mathrm{UV}}  Y_\mathrm{ph}   m_{\mathrm{c}}} \exp{\left(\tau_{\mathrm{r}}\left(s_1, z_1\right)\right)}
\end{equation} 

Here, the exponential factor expresses the attenuation of the UV field within the disk. Analogous to the $\tau_{\mathrm{r}}=1$ constraint for the oxidation case, we use this equation to find the height of the exposed layer $z_\mathrm{1,ph}$, the value of the optical depth at $z_\mathrm{1,ph}$, $\tau_{\mathrm{r, ph}}$ (now generally larger than unity) and the amount of exposed material $\Sigma^{\ast}_\mathrm{ph} = \tau_\mathrm{r, ph}\Phi/\kappa$. Once the layer that is exposed to photolysis is thus characterised, the carbon destruction is calculated using the rates as shown in Eq.~(\ref{eq:sink}). Using Eq.~\ref{app:eq:dS/dt_ph} for the unrestricted photolyis rate, the carbon removal rate then becomes:
\begin{equation}
    \label{eq:sink-ph}
    \left(\frac{d\Sigma_{\mathrm{c}}}{dt}\right)_{\mathrm{ph}} = 
2f_{\mathrm{c}}\cdot \mathrm{min}
\left( 
\frac{\Sigma^{\ast}_{\mathrm{ph}}}{t_{\mathrm{ph}}},  
\frac{\Sigma^{\ast}_{\mathrm{ph}}}{t_{\mathrm{res-ph}}},
\frac{f_{\leq s_1}\Sigma_{\mathrm{tot}}}{t_{\mathrm{mix}}} 
\right)
\end{equation}
where $t_\mathrm{res-ph}$ now follows from the solution to Eq.~(\ref{z1_phot}).
 
\section{Radial dust transport}
\label{app:twopoppy}
We use the \texttt{twopoppy} code by \citet{Birnstiel12, Birnstiel15} to model the radial movement of dust grains. In this code, the dust mass is assigned to two grain sizes, small and large grains. The small grain radius is chosen so that the grains are well coupled to the gas. The radius of the large grains (as well as the ratio of the mass distribution) depends on the local disk conditions, is limited (mainly) by drift and fragmentation and is updated after each timestep. This makes it possible to describe the dust evolution of the disk based on two surface densities. We give here only a short overview over the most important concepts and formulas.

Assuming that the Epstein limit of the drag law applies to all relevant grain sizes in the entire disk, compact spherical grains, a self-similar gas surface density profile and an eddy turn over time of $t_{\mathrm{L}}=1/\Omega$, the Stokes number $\mathrm{St_{mid}}$ of a grain with radius $s$ and specific density $\rho_{\mathrm{s}}$ at the midplane can be written as
\begin{equation}
\label{stokes_til}
\mathrm{St_{mid}}= \frac{s \rho_{\mathrm{s}}}{\Sigma_{\mathrm{g}}}\frac{\pi}{2}\,\;.
\end{equation} 

The small, well-coupled grains have a fixed radius, which we assume to be $s_{\mathrm{s}}=0.1\,\mu$m. The size of the large grains is limited by fragmentation and drift. The size limit due to fragmentation is given by
\begin{equation}
\label{frag_til}
{s_\mathrm{frag}}= f_{\mathrm{f}} \frac{2}{3\pi}\frac{\Sigma_{\mathrm{g}}}{\rho_{\mathrm{s}}\alpha}\frac{v_{\mathrm{f}}^2}{c_{\mathrm{s}}^2}\,\;,
\end{equation} 
where $c_{\mathrm{s}}$ is the isothermal sound speed and $f_{\mathrm{f}}$ is a calibration factor of order unity, leading to grain size slightly below the the fragmentation limit. The size limit due to drift is given by
\begin{equation}
\label{drift_til}
{s_\mathrm{drift}}= f_{\mathrm{d}} \frac{2}{\pi}\frac{\Sigma_{\mathrm{d}}}{\rho_{\mathrm{s}}}\frac{\left(\Omega r\right)^2}{c_{\mathrm{s}}^2}\left|\frac{\mathrm{d}\ln{P}}{\mathrm{d}\ln{r}}\right|^{-1}\,\;,
\end{equation}
where P is the gas pressure and $f_{\mathrm{d}}$ is a calibration factor similar to $f_{\mathrm{f}}$. The radius of the large grains is than chosen as $s_{\mathrm{l}}=\mathrm{min}\left(s_{\mathrm{drift}},s_{\mathrm{frag}} \right)$.

In the fragmentation limited regime, the surface density fraction in large grains is $f_{\mathrm{l}}= 0.75$, and $f_{\mathrm{l}}= 0.97$ in the drift limited regime. We emphasise that these mass distributions and grain sizes are used for the radial and not the vertical dust transport. Only the maximum possible grain size $s_{\mathrm{max}}=s_{l}$ is used in the calculation of $s_1$ and $z_1$ in Section~\ref{sub:calcz1}.

The radial transport of the dust can now be described by solving two advection-diffusion equations:
\begin{equation}
\label{dae_til}
\frac{\partial\Sigma_i}{\partial t} + \frac{1}{r}\frac{\partial}{\partial r}\left[r\left(\Sigma_i v_i - D_i \Sigma_{\rm g} \frac{\partial}{\partial r} 
\left(\frac{\Sigma_i}{\Sigma_{\rm g}}\right)\right)\right]=0
\end{equation}
where $\Sigma_i$ is the surface density of the dust grain, $v_i$ the dust velocity due to drift and gas drag, $D_i$ the diffusivity of the species and the index $i$ refers to the small (s) and the large (l) grains. For the small dust component $v_i$ will be equal to the gas accretion velocity while for the large component the drift velocity and the gas accreation velocity both contribute. Since the Stokes number of a grain is always smaller than unity in this model, the dust diffusivity is assumed to be equal to the gas diffusivity $D_{\mathrm{g}}$ which is considered the same as the gas viscosity.

To follow not only the dust mass and size but also the dust composition, we have modified this code. We use four instead of two types of grains, small and large carbon grains ($s_{\mathrm{c,s}}$, $s_{\mathrm{c,l}}$) and small and large silicate  grains ($s_{\mathrm{s,s}}$, $s_{\mathrm{s,l}}$). As in the original code, $f_{\mathrm{l}}$ is used to distribute the surface density between the particles sizes, and we use $f_{\mathrm{c}}$ to distribute the surface density between the different grain species. The advection-diffusion equation for small carbon grains now includes a sink term and reads:
\begin{equation}
\label{dae_sink}
\frac{\partial\Sigma_{\rm c,s}}{\partial t} + \frac{1}{r}\frac{\partial}{\partial r}\left[r\left(\Sigma_{\rm c,s} v_{\rm s} - D_{\rm s} \Sigma_{\rm g} \frac{\partial}{\partial r} 
\left(\frac{\Sigma_{\rm c,s}}{\Sigma_{\rm g}}\right)\right)\right]= - \frac{d\Sigma_{\rm c}}{dt}
\end{equation}
The advection-diffusion equation for large carbon grains has no sink term:
\begin{equation}
\label{dae_no_sink}
\frac{\partial\Sigma_{\rm c,l}}{\partial t} + \frac{1}{r}\frac{\partial}{\partial r}\left[r\left(\Sigma_{\rm c,l} v_{\rm l} - D_l \Sigma_{\rm g} \frac{\partial}{\partial r} 
\left(\frac{\Sigma_{\rm c,l}}{\Sigma_{\rm g}}\right)\right)\right]= 0
\end{equation}

As in the case of large carbon grains, the advection-diffusion equations for small as well as large silicate grains do not include a sink term.

Once the advection-diffusion equation has been solved, $f_{\mathrm{c}}$ is again calculated, since the radial transport of large and small grains changes the carbon distribution within the disk. By re-calculating a new common carbon fraction for both grain sizes after each timestep, we assume that coagulation and fragmentation have been efficient enough to re-distribute carbon between large and small grains. This maximises the carbon removal efficiency. This carbon fraction value is now the base for the calculation of the carbon destruction. When used in Eq.~(\ref{eq:sink}) and Eq.~(\ref{eq:sink-ph}), it does not indicate the carbon fraction in the exposed layer, but is used to calculate the total amount of carbon available in the exposed layer at the onset of oxidation or photolysis.

Through all these calculations, we assume a constant, self-similar gas surface density profile. The surface density of the destroyed carbon is not added to the gas surface density, but completely removed from the system.

\begin{table*}[htb]
\centering
\begin{tabular}{lp{6cm}}
\hline
symbol & description \\
\hline
\hline
$\Phi$ & disk flaring angle\\
$\Sigma_{\mathrm{c}}$ &  surface density in carbon grains \\ 
$\Sigma_{\mathrm{c,l}}$ &  surface density of large carbon grains\\
$\Sigma_{\mathrm{c,s}}$ &  surface density of small carbon grains\\
$\Sigma_{\mathrm{tot}}$ &  total dust surface density\\
$\Sigma^{\mathrm{\ast}}$ &  dust surface density in exposed layer (oxidation)\\
$\Sigma^{\mathrm{\ast}}_{\mathrm{ph}}$ &  dust surface density in exposed layer (photolysis)\\
$\Omega$ & Keplerian frequency\\
$\alpha$ & turbulence parameter\\
$\epsilon$ & oxygen to gas ratio at $z_{\mathrm{1}}$\\
$\kappa$ & grain opacity\\
$\kappa_0$ & fixed grain opacity at $z_1$\\
$\lambda$ & wavelength\\
$\mu$ & mean molecular weight of neutral hydrogen\\
$\nu_{\mathrm{t}}$ & turbulent viscosity\\
$\rho_{\mathrm{c}}$ & specific carbon grain density\\
$\rho_{\mathrm{d}}$ & disk dust density\\
$\rho_{\mathrm{g}}$ & disk gas density\\
$\rho_{\mathrm{s}}$ & specific grain density\\
$\rho_{\mathrm{tot}}$ & total disk grain density\\
$\sigma$ & grain cross section\\
$\tau_{r}$ & radial optical depth at 0.55\,$\mu$m\\
$\tau_{\mathrm{r, ph}}$ & radial optical depth at 0.55\,$\mu$m at $z_{\mathrm{1,ph}}$\\
$\tau_{z}$ & vertical optical depth at 0.55\,$\mu$m\\
$A$ & parameter for calculation of $Y_{\mathrm{ox}}$\\
$B$ & parameter for calculation of $Y_{\mathrm{ox}}$\\
$D_{\mathrm{g}}$ & gas diffusivity\\
$D_i$ & diffusivity of species i\\
$D_l$ & diffusivity of large grains\\
$D_i$ & diffusivity of small grains\\
$F_{\mathrm{UV}}$ & UV flux\\
$H$ & disk gas scale height\\
$L_{\mathrm{\ast}}$ & stellar luminosity\\
$L_{\mathrm{UV}}$ & stellar luminosity in the UV\\
$M_{\mathrm{\ast}}$ & stellar mass\\
$P$ & gas pressure\\
$T_{\mathrm{g}}$ & gas temperature at $z_{\mathrm{1}}$\\
$T_{\mathrm{i}}$ & gas temperature at $z_{\mathrm{1}}$ at 1\,au\\
St & Stokes number\\
$\mathrm{St}_{\mathrm{mid}}$ & Stokes number in midplane\\
$Y_{\mathrm{ox}}$ & oxidation yield at $z_{\mathrm{1}}$\\
$Y_{\mathrm{ph}}$ & photolysis yield\\
\end{tabular}
\begin{tabular}{lp{6cm}}
$c_{\mathrm{s}}$ & isothermal sound speed\\
$f_{\mathrm{c}}$ & carbon mass fraction in solids\\
$f_{\mathrm{d}}$ & calibration factor for drift limit\\
$f_{\mathrm{f}}$ & calibration factor for fragmentation limit\\
$f_{\mathrm{l}}$ & surface density fraction of large grains\\
$f_{\leq s}$ & mass fraction of grains with radius up to $s$\\
$f_{\leq \mathrm{s_1}}$ & mass fraction of grains with radius up to $s_1$\\
$f_{\geq z}$ & fraction of total surface density above $z$\\
$k_{\mathrm{B}}$ & Boltzmann's constant\\
$h_{\mathrm{gr}}$ & dust scale height \\
$k_{\mathrm{ox}}$ & carbon oxidation rate at $z_{\mathrm{1}}$\\
$m_{\mathrm{gr}}$ & mass of grain with $s_{\mathrm{1}}$\\
$m_{\mathrm{c}}$ & mass of a carbon atom\\
$m_{\mathrm{ox}}$ & mass of an oxygen atom\\
$n_{\mathrm{g}}$ & gas number density at $z_{\mathrm{1}}$\\
$n_{\mathrm{ox}}$ & oxygen number density at $z_{\mathrm{1}}$\\
$p$ & power law index of grain size distribution\\
$q$ & gas temperature power law index at $z_{\mathrm{1}}$\\
$r$ & disk radius\\
$r_{\mathrm{i}}$ & reference radius\\
$s$ & grain radius\\
$s_{\mathrm{c,l}}$ & radius of large carbon grains\\
$s_{\mathrm{c,s}}$ & radius of small carbon grains\\
$s_{\mathrm{drift}}$ & maximum grain radius due to drift\\
$s_{\mathrm{frag}}$ & maximum grain radius due to fragmentation\\
$s_{\mathrm{geo}}$ & transition grain radius from Rayleigh to geometric regime\\
$s_{\mathrm{max}}$ & maximum grain radius at $r$\\
$s_{\mathrm{s,l}}$ & radius of large silicate grains\\
$s_{\mathrm{s,s}}$ & radius of small silicate grains\\
$s_{\mathrm{1}}$ & maximum grain radius at $z_{\mathrm{1}}$\\
$t_{\mathrm{destr}}$ & destruction timescale at $r$\\
$t_{\mathrm{drift}}$ & drift timescale at $r$\\
$t_{\mathrm{L}}$ & eddy turn-over time\\
$t_{\mathrm{mix}}$ & mixing timescale between $z_1$ and midplane\\
$t_{\mathrm{ox}}$ & destruction time of carbon grain with $s_{\mathrm{1}}$ at $z_{\mathrm{1}}$ due to oxidation\\
$t_{\mathrm{res}}$ & residence timescale of grain above $z_1$\\
$t_{\mathrm{res-ph}}$ & residence timescale of grain above $z_{\mathrm{1,ph}}$\\
$v_{i}$ & radial grain velocity\\
$v_{\mathrm{f}}$ & fragmentation velocity\\
$v_{\mathrm{ox}}$ & oxygen thermal velocity at $z_{\mathrm{1}}$\\
$z$ & disk height coordinate\\
$z_{\mathrm{1}}$ & height of the exposed layer (oxidation)\\
$z_{\mathrm{1,ph}}$ & height of the exposed layer (photolysis)\\

\hline
\end{tabular}
\caption{List of notations} 
\label{tab:notations}
\end{table*}
\end{appendix}

\end{document}